\documentclass{IEEEtran}
\IEEEoverridecommandlockouts
\usepackage{cite}
\usepackage{amsmath,amssymb,amsfonts}
\usepackage{algorithmic}
\usepackage{graphicx}
\usepackage{textcomp}
\usepackage{xcolor}
\usepackage{setspace}
\def\BibTeX{{\rm B\kern-.05em{\sc i\kern-.025em b}\kern-.08em
		T\kern-.1667em\lower.7ex\hbox{E}\kern-.125emX}}
\usepackage{subfigure}
\newtheorem{my_theorem}{Theorem}

\newtheorem{my_lemma}{Lemma}

\newtheorem{my_proposition}{Proposition}
\newtheorem{my_remark}{Remark}
\usepackage{enumitem}
\usepackage{array}

\title{\Large{The $\alpha$-$\eta$-$\kappa$-$\mu$ Fading Model: An Exact Statistical Representation }}
\author{Pranay Bhardwaj,~\IEEEmembership{Graduate Student Member,~IEEE}, Eesha Santosh Karnawat,  and S.~M.~Zafaruddin,~\IEEEmembership{Senior Member,~IEEE}
	
	\thanks{The authors are with the Department of Electrical and Electronics Engineering, Birla Institute of Technology and Science, Pilani, Pilani-333031, Rajasthan, India.  Email:\{p20200026, f20190153, syed.zafaruddin\}@pilani.bits-pilani.ac.in.}
	
	\thanks{This work was supported	in part by the Science and Engineering Research Board (SERB), Department of Science and Technology (DST), Government of Indi under Grant MTR/2021/000890.}	
}

\begin{document}
\maketitle 
\begin{abstract}
	The $\alpha$-$\eta$-$\kappa$-$\mu$ is one of the most generalized and flexible channel models having an excellent fit to experimental data from diverse propagation environments. The existing statistical results on the envelope of $\alpha$-$\eta$-$\kappa$-$\mu$ model contain an infinite series, prohibiting its widespread application in the performance analysis of wireless systems. This paper employs a novel approach to derive density and distribution functions of the envelope of the $\alpha$-$\eta$-$\kappa$-$\mu$ fading channel without an infinite series approximation. The derived statistical results are presented using a single Fox's H-function for tractable performance analysis, especially for high-frequency mmWave and terahertz wireless transmissions. We also develop an asymptotic analysis using the Gamma function, which converges to the exact values within a reasonable range of channel parameters. To further substantiate the proposed analysis, we present the exact outage probability and average bit-error-rate (BER) performance of a wireless link subjected to the  $\alpha$-$\eta$-$\kappa$-$\mu$ fading model using a single trivariate Fox's H-function. We obtain the diversity order of the system by analyzing the outage probability at a high signal-to-noise (SNR) ratio. We use simulations to demonstrate the significance of the developed statistical results compared with the existing infinite series representation of the $\alpha$-$\eta$-$\kappa$-$\mu$ model.
	\end{abstract}			

\section{Introduction}
The sixth generation (6G) and beyond wireless communication systems are expected to utilize high-frequency spectrum bands such as millimeter-wave (mmWave), and terahertz (THz) \cite{Dang_2020_nature,Akyildiz_2020}. Statistical characterization and modeling of mmWave and THz channels are necessary for a practical design and deployment of next-generation wireless systems \cite{Serghiou_2022_THz_survey}. A desirable statistical model should fit the measurement data in most situations and be mathematically tractable for performance analysis. Initial research proposed classical Rayleigh and Rice distributions for small-scale fading amplitudes for the mmWave band, specifically at a carrier frequency of $28$ \mbox{GHz} and $60$ \mbox{GHz} \cite{Thomas1994,Smulders2009,Moon2005,Samimi2016}. However, single-parameter models such as Rayleigh, Rice, Hoyt, Nakagami-m, and Weibull may not provide enough flexibility to accurately  fit the measurement data in some  propagation scenarios, requiring more generalized and flexible models. 

Based on recent small-scale fading measurements of the 28 GHz outdoor millimeter-wave channels \cite{Samimi2016},  the fluctuating two-ray (FTR) fading model has been proposed as a versatile model that can provide a much better fit than the Rician fading model \cite{Romero2016}. The authors in \cite{Marins2019_alpha_eta_kapp_mu} examined the applicability of  Rayleigh, Rice, $\alpha$-$\mu$, $\kappa$-$\mu$, $\eta$-$\mu$,  and $\alpha$-$\eta$-$\kappa$-$\mu$  for mmWave transmission in the range from $55$ \mbox{GHz} to $65$ \mbox {GHz} thorough extensive measurement campaign in an indoor environment. They found that $\alpha$-$\mu$, $\kappa$-$\mu$, and $\eta$-$\mu$ can be applied to most situations. However, these models fail to provide a good fit in some of the intricate occasions even in the lower frequency range ($55$ \mbox{GHz} to $65$ \mbox {GHz}) and thus expected to fail at a much higher mmWave/THz frequency in most of the propagation scenarios. Further, \cite{Marins2019_alpha_eta_kapp_mu} demonstrated that the $\alpha$-$\eta$-$\kappa$-$\mu$  is the best option fitting over a wide range of situations, including intricate and rare events.

Recently, the authors in \cite{Papasotiriou2021_scientific_report} conducted extensive experiments in indoor environments to model short-term fading of THz signal at a frequency of $143$ GHz. It is shown in \cite{Papasotiriou2021_scientific_report} that classical fading models (Rayleigh, Rice, and Nakagami-m) lack fitting accuracy, whereas the $\alpha$-$\mu$  distribution provides an excellent fit except in few cases where the experimental probability distribution function (PDF) becomes bimodal. The authors in \cite{Du2022_FTR_THz_RIS} used the actual measurement data demonstrating the better fit of the FTR model than the $\alpha$-$\mu$  at a frequency of $304.2$ \mbox{GHz}. However, the FTR model contains an infinite sum to generate the PDF and cumulative distribution function (CDF). Further, it is unclear how the $\alpha$-$\mu$ and FTR models would perform over a true THz frequency (for example, at $1$ THz). It can be anticipated that more generalized and flexible fading models would be required to accurately model the mmWave and THz signals at higher frequencies under diverse propagation scenarios. 
 
In the seminal work \cite{Yacoub_2016_alpha_eta_kappa_mu}, M. D. Yacoub proposed the $\alpha$-$\eta$-$\kappa$-$\mu$ model as the most flexible and comprehensive wireless channel fading model. Later, Silva \emph{et al.} \cite{Silva_2020_alpha_eta_kappa_mu} improved the mathematical representation of the PDF and CDF of the channel envelope. The $\alpha$-$\eta$-$\kappa$-$\mu$ model encapsulates various fading characteristics such as nonlinearity of the propagation medium, number of multi-path clusters, scattering level, and power of the dominant components. The multi-parameter $\alpha$-$\eta$-$\kappa$-$\mu$ model encompasses $ \alpha $-$ \mu $, $ \kappa $-$ \mu $, $ \eta$-$ \mu $, $ \alpha $-$\eta$-$ \mu $, and $ \alpha $-$\kappa$-$ \mu $  as the special cases, and thus can fit the experimental data for mmWave and THz signals more accurately in a variety of scenarios. However, the statistical results of the $\alpha$-$\eta$-$\kappa$-$\mu$ model presented in \cite{Yacoub_2016_alpha_eta_kappa_mu} \cite{Silva_2020_alpha_eta_kappa_mu} contains an infinite series comprising of mathematical functions such as regularized hypergeometric function and generalized Laguerre polynomial. These functions have  complex mathematical formulations and do not present a straightforward insight into propagation channel behavior.

Further, the asymptotic expressions of the PDF and CDF of the channel envelope require extreme values for convergence, limiting its usage in performance analysis. Thus, analytical complexity involving infinite series containing complicated functions has prohibited a widespread application of the $\alpha$-$\eta$-$\kappa$-$\mu$ model for performance analysis despite its generality and flexibility compared with other fading models. Nevertheless, there are research works using the  $\alpha$-$\eta$-$\kappa$-$\mu$ model for wireless communications \cite{Li_2017_alpha_eta_kappa_mu,Mathur_2018_alpha_eta_kappa_mu,Jia2018_alpha_eta_kapp_mu,Moualeu_2019_alpha_eta_kappa_mu,Singh2019_alpha_eta_kapp_mu,Ai_2020_alpha_eta_kappa_mu,Al2020_alpha_eta_kapp_mu,Saraereh2020_alpha_eta_kapp_mu,Soni2020_alpha_eta_kapp_mu,Vishwakarma2021_alpha_eta_kapp_mu,GOSWAMI2019_alpha_eta_kapp_mu,Kavaiya2020_alpha_eta_kapp_mu}. It should be mentioned that most of the existing literature \cite{Li_2017_alpha_eta_kappa_mu,Mathur_2018_alpha_eta_kappa_mu,Jia2018_alpha_eta_kapp_mu,Moualeu_2019_alpha_eta_kappa_mu,Singh2019_alpha_eta_kapp_mu,Ai_2020_alpha_eta_kappa_mu,Al2020_alpha_eta_kapp_mu,Saraereh2020_alpha_eta_kapp_mu,Soni2020_alpha_eta_kapp_mu,Vishwakarma2021_alpha_eta_kapp_mu} presented  analysis of performance metrics such outage probability, average bit-error rate, and ergodic capacity   using Meijer's G-function and/or Fox's H-function. Infinite-series expansion in the PDF and CDF (with special functions)  can be avoided, provided the performance analysis ultimately involves Fox's H-function.

In this paper, we employ a novel approach to derive an exact statistical representation of the $\alpha$-$\eta$-$\kappa$-$\mu$ fading model for mmWave and THz wireless systems. We present the PDF and CDF of the channel envelope using a single Fox's-H function without any infinite series approximation. Note that the use of  Fox's H functions is widespread in the research fraternity to analyze the performance of wireless systems over generalized fading models. Further, we develop an asymptotic analysis on the envelope of the $\alpha$-$\eta$-$\kappa$-$\mu$ using a more straightforward Gamma function converging to the exact with reasonable values of channel parameters. We demonstrate the superiority of the developed statistical results by analyzing the outage probability and average bit-error-rate (BER) performance of a wireless link subjected to the  $\alpha$-$\eta$-$\kappa$-$\mu$ fading model.
\linespread{0.90}
\section{A Primer on  $\alpha$-$\eta$-$\kappa$-$\mu$ Fading Model}
In this section, we revisit the $ \alpha $-$ \eta $-$ \kappa $-$ \mu $ model, as described in \cite{Yacoub_2016_alpha_eta_kappa_mu}. This model includes almost all short-term propagation phenomena  to generalize the fading model for a wireless channel. The envelope $R$ of the $ \alpha $-$ \eta $-$ \kappa $-$ \mu $  is given by  \cite{Yacoub_2016_alpha_eta_kappa_mu}\cite{Silva_2020_alpha_eta_kappa_mu}
\begin{flalign} \label{eq:Ralpha}
R^\alpha = \sum_{i=1}^{\mu_x} (X_i+{\lambda_{x_i}})^2 + \sum_{i=1}^{\mu_y} (Y_i+{\lambda_{y_i}})^2
\end{flalign}
where $\alpha$ denotes the non-linearity of the channel, $\mu_x$ and $\mu_y$ denote the number of multi-path clusters of in-phase component and quadrature component, respectively, $\lambda_{x_i}$ and $\lambda_{y_i}$ are the average values of the in-phase and quadrature components of the multi-path waves of the $i$-th cluster, respectively, and $X_i\sim {\cal{N}} (0, \sigma_x^2)$ and $Y_i\sim {\cal{N}} (0,\sigma_y^2)$ are mutually independent Gaussian processes, where $\sigma_x^2$ and $\sigma_y^2$ are variances of in-phase and quadrature components of the multi-path waves, respectively. In general, the $ \alpha $-$ \eta $-$ \kappa $-$ \mu $ model is quantified by seven different parameters, namely $\alpha$, $\eta$, $\kappa$, $\mu$, $p$, $q$, and $\hat{r}$. To define these parameter, denote the power of in-phase $(x)$ and quadrature-phase $(y)$ components of dominant $(d)$ waves and scattered $ (s) $ waves as $P_{ab}$, where $a \in \{d,s\}$ and $b \in \{x,y\}$. Thus, we define the parameters as $\eta = \frac{P_{sx}}{P_{sy}}$,  $\kappa = \frac{P_{dx} + P_{dy}}{P_{sx} + P_{sy}}$, $\mu = \frac{\mu_x+\mu_y}{2}$, $p = \frac{\mu_x}{\mu_y}$, $q = \frac{P_{dx}}{P_{dy}}/\frac{P_{sx}}{P_{sy}}$, and $\hat{r}= \sqrt[\alpha]{\mathbb{E}[R^{\alpha}]}$.

For statistical analysis of the wireless systems, the density and distribution functions of the channel envelope are required. Using \eqref{eq:Ralpha}, the authors in \cite{Silva_2020_alpha_eta_kappa_mu} presented the PDF $f_R(r)$ of the envelope $R$ in terms of the generalized Laguerre polynomial ($L_n$) and  the regularized hypergeometric function (${}_0\tilde{F}_1 $):
\begin{flalign} \label{eq:pdf_series}
	f_R(r) &= \frac{\alpha (\xi \mu)^\mu}{\exp\Big(\frac{(1+pq)\kappa\mu}{\delta}\Big)} \bigg(\frac{p}{\eta}\bigg)^{\frac{p\mu}{1+p}} \frac{r^{\alpha\mu-1}}{\hat{r}^{\alpha \mu} } \exp\Big(-\frac{r^\alpha p \xi \mu}{\hat{r}^\alpha \eta}\Big) \nonumber \\ & \times \sum_{n=0}^{\infty} \Big(\frac{r^\alpha \xi \mu (p-\eta)}{\hat{r}^\alpha \eta}\Big)^n L_n^{\frac{\mu}{1+p}-1} \Big(\frac{\eta \kappa \mu}{\delta (\eta-p)}\Big) \nonumber \\ & \times {}_0\tilde{F}_1 \Big(;\mu+n;\frac{p^2qr^\alpha \kappa \xi \mu^2}{\hat{r}^\alpha \delta \eta}\Big) 
\end{flalign}
where $\xi = \frac{(1+\eta)(1+\kappa)}{(1+p)}$ and $\delta = \frac{(1+q\eta)(1+p)}{(1+\eta)}$. It can be seen that the PDF $f_R(r)$ contains an infinite series representation, which approximates the system performance when a finite number of terms are used for the convergence of the distribution function. It is always desirable to express the statistics of the channel envelope in an exact form using tractable mathematical functions for efficient performance analysis and numerical computations.

\section{Exact Statistical Derivation of $\alpha$-$\eta$-$\kappa$-$\mu$  Model}
In this section, we derive exact expressions of the PDF and CDF of the channel envelope distributed according to the $\alpha$-$\eta$-$\kappa$-$\mu$ model using a single Fox's H-function. We also develop asymptotic expressions for PDF and CDF in terms of simple algebraic functions. We denote the multi-variate Fox's H-function as given in \cite [A.1] {Mathai_2010}. We define $ \psi_1 = \frac{p\alpha\mu^{2}\xi^{1+\frac{\mu}{2}}\delta^{\frac{\mu}{2}-1}q^{\frac{1+p-p\mu}{2+2p}}\eta^{-\frac{1+p+p\mu}{2+2p}}}{\kappa^{\frac{\mu}{2}-1}\exp \left (\frac{(1+pq)\kappa\mu}{\delta}\right)}$, $ \psi_2$ = $\alpha-1$, $ \psi_3= \frac{p\xi\mu}{\eta\hat{r}^\alpha} $, $ A_{1} = \frac{p\mu}{1+p}$-$1, A_{2}= \frac{\mu}{1+p}$-$1, A_{3}= \frac{(\eta-p)\xi\mu}{\eta\hat{r}^\alpha} $, $ A_{4} = 2p\mu \sqrt{\frac{q\kappa\xi}{\eta\delta\hat{r}^\alpha}} $, and $A_{5} = 2\mu\sqrt{\frac{\kappa\xi}{\delta\hat{r}^\alpha}} $. 

First, we use \eqref{eq:Ralpha} to present the PDF of the channel envelope in the following theorem:
\begin{my_theorem}
The PDF of the channel envelope for the  $\alpha$-$\eta$-$\kappa$-$\mu$ fading model is given by
\begin{flalign} \label{eq:pdf_new_without_pe}
	&f_{R}(r) = \frac{\psi_1 \pi^2 2^{(2-\mu)} A_4^{A_1} A_5^{A_2} r^{\alpha\mu-1} e^{-\psi_3 r^\alpha}}{(\hat{r}^\alpha)^{1+\frac{\mu}{2}}} \nonumber \\ &\times H^{0,1;1,0;1,1;1,0}_{1,1;0,1;2,3;1,3} \Bigg[ \begin{matrix}~ V_1~ \\ ~V_2~ \end{matrix} \Bigg| A_{3}r^{\alpha}, \frac{A_4^2}{4}r^{\alpha},\frac{A_5^2}{4}r^{\alpha}\Bigg],
\end{flalign}	
where $V_1 = \big\{(-A_2;1,0,1)\big\}: \big\{(-,-)\big\} ; \big\{(-A_1,1)(\frac{1}{2},1)\big\} ;\big\{(\frac{1}{2},1)\big\} $ and $V_2 = \big\{(-1-A_1 -A_2;1,1,1) \big\} : \big\{(0,1) \big\} ; \big\{(0,1),(-A_1,1),(\frac{1}{2},1) \big\} ; \big\{(0,1),(-A_2,1),(\frac{1}{2},1) \big\}$.
\end{my_theorem}

\begin{IEEEproof}	
See Appendix A.
\end{IEEEproof}

It can be seen that the PDF of \eqref{eq:pdf_new_without_pe} contains a single tri-variate Fox's H-function without any infinite series, as compared to \eqref{eq:pdf_series}. It should be noted that previous studies have used Fox's H-function equivalent of the mathematical functions present in \eqref{eq:pdf_new_without_pe} to directly express its PDF in terms of the product of Fox's H-functions without removing the infinite series for performance analysis. Recently,  Fox's H functions are finding applications in wireless communication research unifying performance analysis for intricate  fading distributions.
\begin{my_remark}
For a fair comparison between the computational complexity of the existing formulation presented in equation \eqref{eq:pdf_series} with the proposed solution in equation \eqref{eq:pdf_new_without_pe}, we rely on a qualitative comparison since there is no optimized computational routine available in commonly used software such as MATLAB and MATHEMATICA for the tri-variate Fox's H-function, as opposed to the mathematical functions used in \eqref{eq:pdf_series}. Let the computational complexity of a single term in \eqref{eq:pdf_series} be denoted as $C^{\rm series}$ and the complexity of \eqref{eq:pdf_new_without_pe} as $C^{\rm fox}$. Although it may be true that $C^{\rm series}<C^{\rm fox}$, as the value of $n$ increases, the overall complexity $nC^{\rm series}$ may eventually exceed $C^{\rm fox}$ for a reasonable number of summands $n$.
\end{my_remark}

Furthermore,   the asymptotic expansion  of  the Fox's H-function may have better characteristics providing more accurate approximation over a wide range of parameters:
\begin{my_proposition}
An asymptotic expression for PDF of the $\alpha$-$\eta$-$\kappa$-$\mu$ fading channel is given by 
\begin{flalign} \label{eq:asympt_pdf}
	\lim_{r\to0}f_{R}(r) = & \frac{\psi_1 \pi^2 2^{(2-\mu)} A_4^{A_1} A_5^{A_2} r^{\alpha\mu-1} }{4(\hat{r}^\alpha)^{1+\frac{\mu}{2}} e^{\psi_3 r^\alpha}} \frac{1}{\Gamma(\mu) \Gamma(1+A_2) \Gamma(\frac{1}{2})}
\end{flalign}
\end{my_proposition}

\begin{IEEEproof}
The existing literature derives the asymptotic expression by computing the residue of dominant pole of a multi-variate Fox's H-function, which sometimes become tedious. We convert the tri-variate Fox's H-function using the  product of three Meijer's G-function to derive the asymptotic expression of the PDF  using simpler mathematical functions. Thus,  eliminating the linear combination terms of the tri-variate Fox's H-function, we can express the   PDF in \eqref{eq:pdf_new_without_pe} as
		\begin{flalign}\label{eq:three_meijer}
			f_{R}(r) \approx & \frac{\psi_1 \pi^2 2^{(2-\mu)} A_4^{A_1} A_5^{A_2} r^{\alpha\mu-1} e^{-\psi_3 r^\alpha}}{(\hat{r}^\alpha)^{1+\frac{\mu}{2}}} G^{1,0}_{0,1}\bigg(\begin{matrix} - \\ 0 \end{matrix} \bigg| A_3 r^\alpha \bigg) \nonumber \\ &  G^{1,1}_{2,4}\bigg(\begin{matrix} -A1, \frac{1}{2} \\ 0, -A_1, \frac{1}{2}, -1-A_1-A_2 \end{matrix} \bigg| \frac{A_4^2 r^\alpha}{4} \bigg)  \nonumber \\ &   G^{1,0}_{1,3}\bigg(\begin{matrix} \frac{1}{2} \\ 0, -A_2, \frac{1}{2} \end{matrix} \bigg| \frac{A_5^2 r^\alpha}{4} \bigg)
	\end{flalign}

		Applying \cite{Meijers_asymp_low_snr}  for the asymptotic expansion at $r \to 0$ for Meijer's G-functions in \eqref{eq:three_meijer}, leads to the asymptotic PDF in \eqref{eq:asympt_pdf}.
\end{IEEEproof}
It should be mentioned that the authors in \cite{Silva_2020_alpha_eta_kappa_mu} provided  asymptotic expression for the PDF of the  $\alpha$-$\eta$-$\kappa$-$\mu$ distribution without any series expression.  However, as demonstrated through simulations in the next Section,  the proposed asymptotic PDF in  \eqref{eq:asympt_pdf} offers a more accurate  representation than \cite{Silva_2020_alpha_eta_kappa_mu}.

In the following lemma, we derive the CDF of the  $\alpha$-$\eta$-$\kappa$-$\mu$ fading channel.
\begin{my_lemma}
	The CDF of the channel envelope for the $\alpha$-$\eta$-$\kappa$-$\mu$ fading model is given by		
	\begin{flalign} \label{eq:cdf_new_without_pe}
		&F_{R}(r) = \frac{\psi_1 \pi^2 2^{(2-\mu)} A_4^{A_1} A_5^{A_2} r^{{\alpha\mu}}}{(\hat{r}^\alpha)^{1+\frac{\mu}{2}}} \nonumber \\ &\times H^{0,2;1,0;1,1;1,0;1,0}_{2,2;0,1;2,3;1,3;0,1} \Bigg[\begin{matrix}~ V_3~ \\ ~V_4~ \end{matrix} \Bigg| A_{3}r^{\alpha}, \frac{A_4^2}{4}r^{\alpha},\frac{A_5^2}{4}r^{\alpha}, \psi_3 r^\alpha \Bigg],
	\end{flalign}	
	where $V_3 = \big\{(-A_2;1,0,1,0),(1-{\alpha\mu}; \alpha,\alpha,\alpha,\alpha)\big\}: \big\{(-,-)\big\} ; \big\{(-A_1,1)(\frac{1}{2},1)\big\} ; \big\{(\frac{1}{2},1)\big\} ; \big\{(-,-)\big\} $ and $V_4 = \big\{(-1-A_1-A_2;1,1,1,0),(-{\alpha\mu};\alpha,\alpha,\alpha,\alpha) \big\} : \big\{(0,1) \big\} ; \big\{(0,1),(-A_1,1),(\frac{1}{2},1) \big\} ; \big\{(0,1),(-A_2,1),(\frac{1}{2},1) \big\} ; \\ \big\{(0,1)\big\}$.
\end{my_lemma}

\begin{IEEEproof}
See Appendix B.
\end{IEEEproof}
Note that unifying the statistical representation for the PDF and CDF using a single Fox's H-function can facilitate a better tractability of performance analysis.In the next section, we use the derived statistical results of Theorem 1 and Lemma 1 to analyze  the performance  of a wireless link subjected to $\alpha$-$\eta$-$\kappa$-$\mu$ short-term fading. 

\section{Performance Analysis of a Communication Link Over $\alpha$-$\eta$-$\kappa$-$\mu$ Fading }
Consider a source that transmits the information to the destination using a single antenna. We define the signal-to-noise ratio (SNR)  at the receiver as $\gamma = \bar{\gamma}|R|^2$, where $\bar{\gamma}$ is the average SNR. We require the PDF and CDF of the SNR $\gamma$ for statistical performance analysis. Using standard transformation of random variable, the PDF of SNR is given as 
$f_\gamma(\gamma) = \frac{1}{2\sqrt{\gamma \bar{\gamma}}}f_R(\sqrt{\frac{\gamma}{\bar{\gamma}}})$ and $F_\gamma(\gamma) = F_R(\sqrt{\frac{\gamma}{\bar{\gamma}}})$, where $f_R(r)$ is given in \eqref{eq:pdf_new_without_pe} and $F_R(r)$ is given in \eqref{eq:cdf_new_without_pe}.

In the following subsections, we use the derived statistical results to present the exact analysis of the outage probability and average BER of a wireless link subjected to $\alpha$-$\eta$-$\kappa$-$\mu$ fading.

\begin{figure*}[t]
	\centering
	\subfigure[PDF]{\includegraphics[scale=0.31]{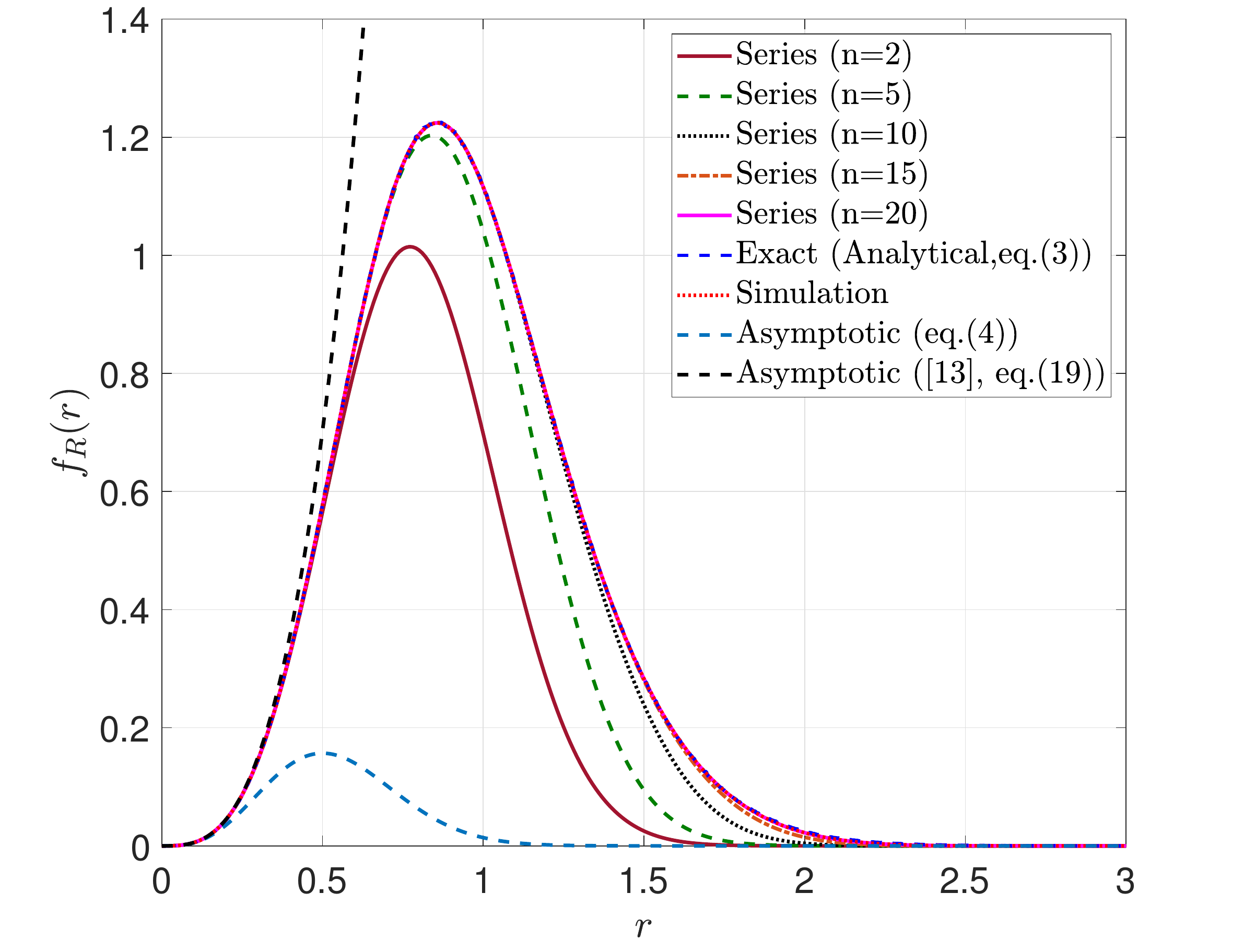}} 
	\subfigure[CDF]{\includegraphics[scale=0.31]{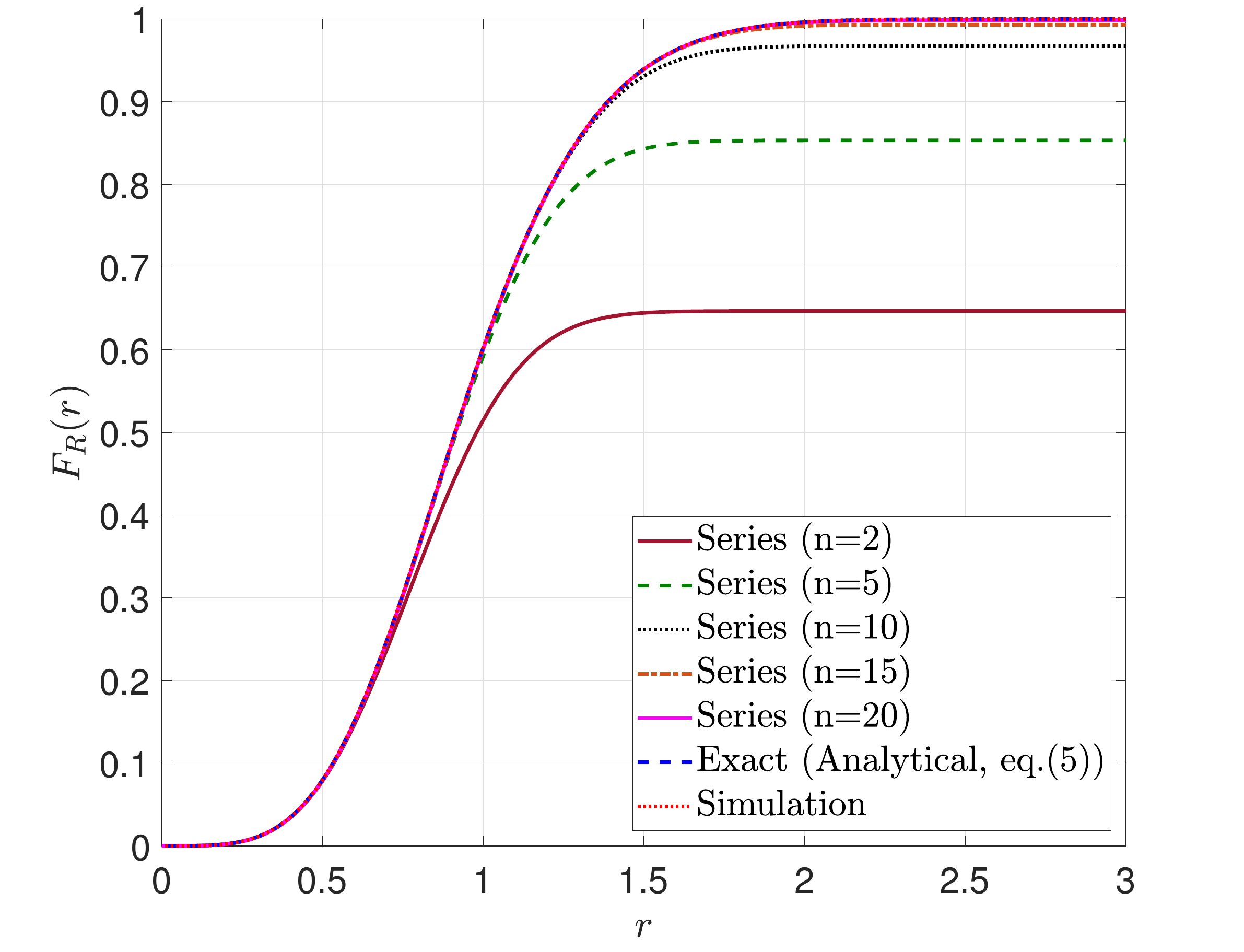}} 
	\caption{Density and distribution functions of the envelope for  $\alpha$-$\eta$-$\kappa$-$\mu$ fading channel: comparison between the derived exact  and infinite-series \cite{Silva_2020_alpha_eta_kappa_mu}.}
	\label{fig:pdf_cdf}	
\end{figure*}

\subsection{Outage Probability}
The outage probability is defined as the probability of instantaneous SNR being less than a threshold SNR value $\gamma_{\rm th}$ i.e.,  $P_{\rm out} = Pr(\gamma < \gamma_{\rm th}) = F_{\gamma}(\gamma_{\rm th})$. Using \eqref{eq:cdf_new_without_pe} in $F_\gamma(\gamma_{\rm th}) = F_R(\sqrt{\frac{\gamma_{\rm th}}{\bar{\gamma}}})$, the outage probability can be expressed as
\begin{flalign} \label{eq:outage_new_without_pe}
	&P_{\rm out} = \frac{\psi_1  \pi^2 2^{(2-\mu)} A_4^{A_1} A_5^{A_2} \gamma_{\rm th}^{\frac{\alpha\mu}{2}}}{(\hat{r}^\alpha)^{1+\frac{\mu}{2}} \bar{\gamma}^{\frac{\alpha\mu}{2}}} \nonumber \\ &\times H^{0,2;1,0;1,1;1,0;1,0}_{2,2;0,1;2,3;1,3;0,1} \Bigg[\begin{matrix}~ V_3~ \\ ~V_4~ \end{matrix} \Bigg| \frac{A_{3}\gamma_{\rm th}^{\frac{\alpha}{2}}}{\bar{\gamma}^{\frac{\alpha}{2}}}, \frac{{A_{4}^{2}}\gamma_{\rm th}^{\frac{\alpha}{2}}}{4\bar{\gamma}^{\frac{\alpha}{2}}},\frac{{A_{5}^{2}}\gamma_{\rm th}^{\frac{\alpha}{2}}}{4 \bar{\gamma}^{\frac{\alpha}{2}}}, \frac{\psi_3 \gamma_{\rm th}^{\frac{\alpha}{2}}}{\bar{\gamma}^{\frac{\alpha}{2}}} \Bigg],
\end{flalign}	
The  outage probability in a high SNR regime can provide useful insight for system design. We compute the residue at dominant poles to derive an  asymptotic expression for the outage probability:
\begin{flalign}
	P_{\rm out}^\infty =& \frac{\psi_1 2^{(2-\mu)} A_4^{A_1} A_5^{A_2} \gamma_{\rm th}^{\frac{\alpha\mu}{2}}} { (\hat{r}^\alpha)^{1+\frac{\mu}{2}} \bar{\gamma}^{\frac{\alpha\mu}{2}}  \alpha \mu} \frac{1}{\Gamma(\mu) \Gamma(A_2)} .
\end{flalign}
The asymptotic outage can be represented as $P_{\rm out}^\infty = G_c \bar{\gamma}^{G_d}$, where $G_c$ and $G_d$ denote  coding gain and diversity order, respectively. Observing the exponent of average SNR $\bar{\gamma}$, we can deduce that the diversity order $G_d$ of the system over the $\alpha$-$\eta$-$\kappa$-$\mu$ fading channel is $\frac{\alpha\mu}{2}$. Thus, the other parameters, such as $\eta$ and $\kappa$, affect the system's coding gain $G_c$.

\subsection{Average BER}
Average BER for binary modulations can be expressed using the CDF \cite{Ansari2011}:
\begin{flalign} \label{eq:ber_eqn}
	\overline{P_e} = \frac{ q_{\rm m}^{p_{\rm m}}}{2\Gamma(p_m)} \int_{0}^{\infty} \gamma^{p_m-1} e^{-q_m\gamma} F_\gamma(\gamma) d\gamma
\end{flalign}
where $p_m$ and $q_m$ determine the type of modulation scheme used. Thus, binary phase shift keying (BPSK) can be represented by $\{ p_m  = 0.5, q_m  = 1\}$, while differential phase shift keying (DPSK) and binary frequency shift keying (BFSK) are characterized by $ \{p_m = 1, q_m = 1 \}$, and $ \{p_m = 0.5, q_m = 0.5\} $, respectively.

Substituting the CDF of \eqref{eq:cdf_new_without_pe} in \eqref{eq:ber_eqn}, using the integral form of Fox's H-function, and changing the order of integration, we get
\small
\begin{flalign} \label{eq:ber_derive}
	& \overline{P_e} =  \frac{\psi_1  \pi^2 2^{(2-\mu)} A_4^{A_1} A_5^{A_2}  q_m^{p_m}}{2\Gamma(p_m)\bar{\gamma}^{\frac{\alpha\mu}{2}}} \Big(\frac{1}{2\pi i}\Big)^{4} \nonumber 
	 \\ & \times \int_{\mathcal{L}_1}\int_{\mathcal{L}_2}\int_{\mathcal{L}_3} \int_{\mathcal{L}_4} \frac{\Gamma(-s_1) \Gamma(-s_2)\Gamma(1+{A_1}+s_2)}{\Gamma(1+A_1+s_2)\Gamma(\frac{1}{2}+s_2)\Gamma(\frac{1}{2}-s_2)}
	\nonumber \\ &  \times \frac{\Gamma(-s_3)}{\Gamma(1+A_2+s_3)\Gamma(1/2+s_3)\Gamma(1/2-s_3)} \nonumber \\ & \times
	\frac{\Gamma(1+{A_2}+s_1+s_3)}{\Gamma(2+{A_2}+s_1+s_3+{A_1}+s_2)} \nonumber \\ & \times \frac{\Gamma(\alpha(s_1+s_2+s_3+s_4)+{\alpha\mu})}{\Gamma(\alpha(s_1+s_2+s_3+s_4)+{\alpha\mu}+1)} \nonumber \\ & \times
	\Big(\frac{A_{3}}{\bar{\gamma}^\frac{\alpha}{2}}\Big)^{s_1} \Big(\frac{{A_{4}^{2}}}{4\bar{\gamma}^{\frac{\alpha}{2}}}\Big)^{s_2}   \Big(\frac{{A_{5}^{2}}}{4\bar{\gamma}^{\frac{\alpha}{2}}}\Big)^{s_3} \Big(\frac{\psi_3}{\bar{\gamma}^\frac{\alpha}{2}}\Big) ds_1 ds_2 ds_3 ds_4 \times I_3
\end{flalign}
\normalsize
where $\mathcal{L}_1$, $\mathcal{L}_2$, and $\mathcal{L}_3$, and $\mathcal{L}_4$ denote the contour in the complex plane. The inner integral $I_3$ can be simplified using \cite[(3.381/4)]{Gradshteyn} to
\begin{flalign}
	I_3 &= \int_{0}^{\infty} \gamma^{p_m-1} e^{-q_m\gamma} \gamma^{\frac{\alpha}{2}(s_1+s_2+s_3+s_4)+\frac{\alpha\mu}{2}}  d\gamma \nonumber 
	 \\ &   = q_m^{-(\frac{\alpha}{2}(s_1+s_2+s_3+s_4)+\frac{\alpha\mu}{2})-p_m} \nonumber \\ &  \times \Gamma(\frac{\alpha}{2}(s_1+s_2+s_3+s_4)+\frac{\alpha\mu}{2}+p_m)
\end{flalign}
Substituting $I_3$ in \eqref{eq:ber_derive} and applying the definition of multivariate Fox's H-function, we get the average BER for the considered system as
\begin{flalign} \label{eq:ber_new_without_pe}
	&\overline{P_e} = \frac{\psi_1 \pi^2 2^{(2-\mu)} A_4^{A_1} A_5^{A_2} }{2\Gamma(p_m) \bar{\gamma}^{\frac{\alpha\mu}{2}} q_m^{\frac{\alpha\mu}{2}}}\nonumber \\ & H^{0,3;1,0;1,1;1,0;1,0}_{3,2;0,1;2,3;1,3;0,1} \Bigg[\begin{matrix} ~V_3~ \\ ~V_4~ \end{matrix} \Bigg| \frac{A_{3}}{\bar{\gamma}^{\frac{\alpha}{2}} q_m^{\frac{\alpha}{2}}}, \frac{{A_{4}^{2}}\gamma^{\frac{\alpha}{2}}}{4\bar{\gamma}^{\frac{\alpha}{2}} q_m^{\frac{\alpha}{2}}},\frac{{A_{5}^{2}}\gamma^{\frac{\alpha}{2}}}{4\bar{\gamma}^{\frac{\alpha}{2}} q_m^{\frac{\alpha}{2}}}, \frac{\psi_3}{\bar{\gamma}^{\frac{\alpha}{2}} q_m^{\frac{\alpha}{2}}} \Bigg],
\end{flalign}	
where $V_3 = \big\{(-{A_2};1,0,1,0),(1-{\alpha\mu}; \alpha,\alpha,\alpha,\alpha), (1-\frac{\alpha\mu}{2}-p_m; \frac{\alpha}{2},\frac{\alpha}{2},\frac{\alpha}{2},\frac{\alpha}{2})\big\}: \big\{(-,-)\big\} ; \big\{(-{A_1},1)(\frac{1}{2},1)\big\} ; \big\{(\frac{1}{2},1)\big\} ; \big\{(-,-)\big\} $ and $V_4 = \big\{(-1-{A_1} -{A_2};1,1,1,0),(-\frac{\alpha\mu};\alpha,\alpha,\alpha,\alpha) \big\} : \big\{(0,1) \big\} ; \big\{(0,1),(-A_1,1),(\frac{1}{2},1) \big\} ; \big\{(0,1),(-A_2,1),(\frac{1}{2},1) \big\} ; \\ \big\{(0,1)\big\}$.

\begin{figure*}[t]
	\centering
	\subfigure[Outage Probability, $p=1$, $q=1$, $\eta=1.01$]{\includegraphics[scale=0.32]{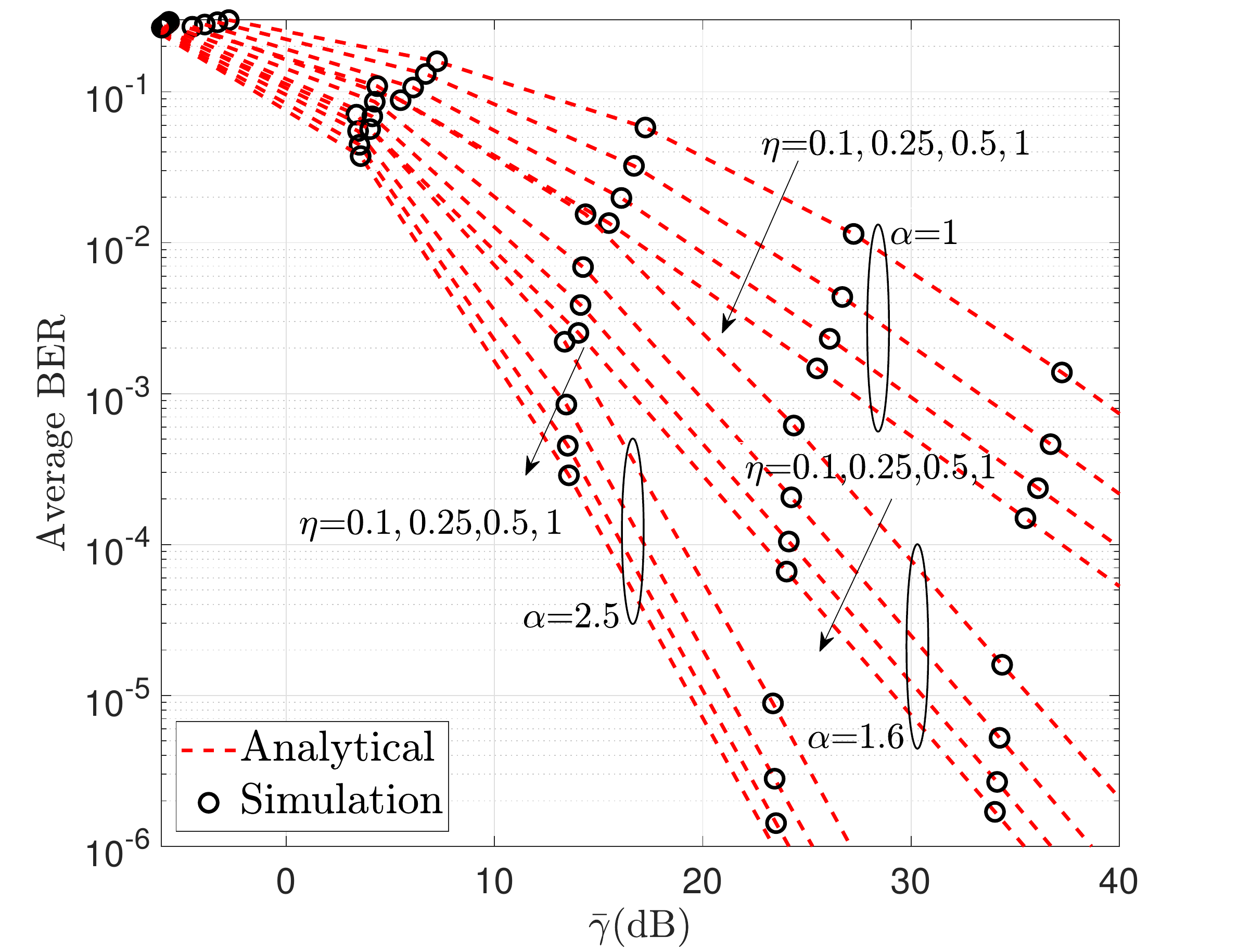}} 
	\subfigure[Average BER, $p=3$, $q=1$, $\mu=2$, $\kappa=1$]{\includegraphics[scale=0.32]{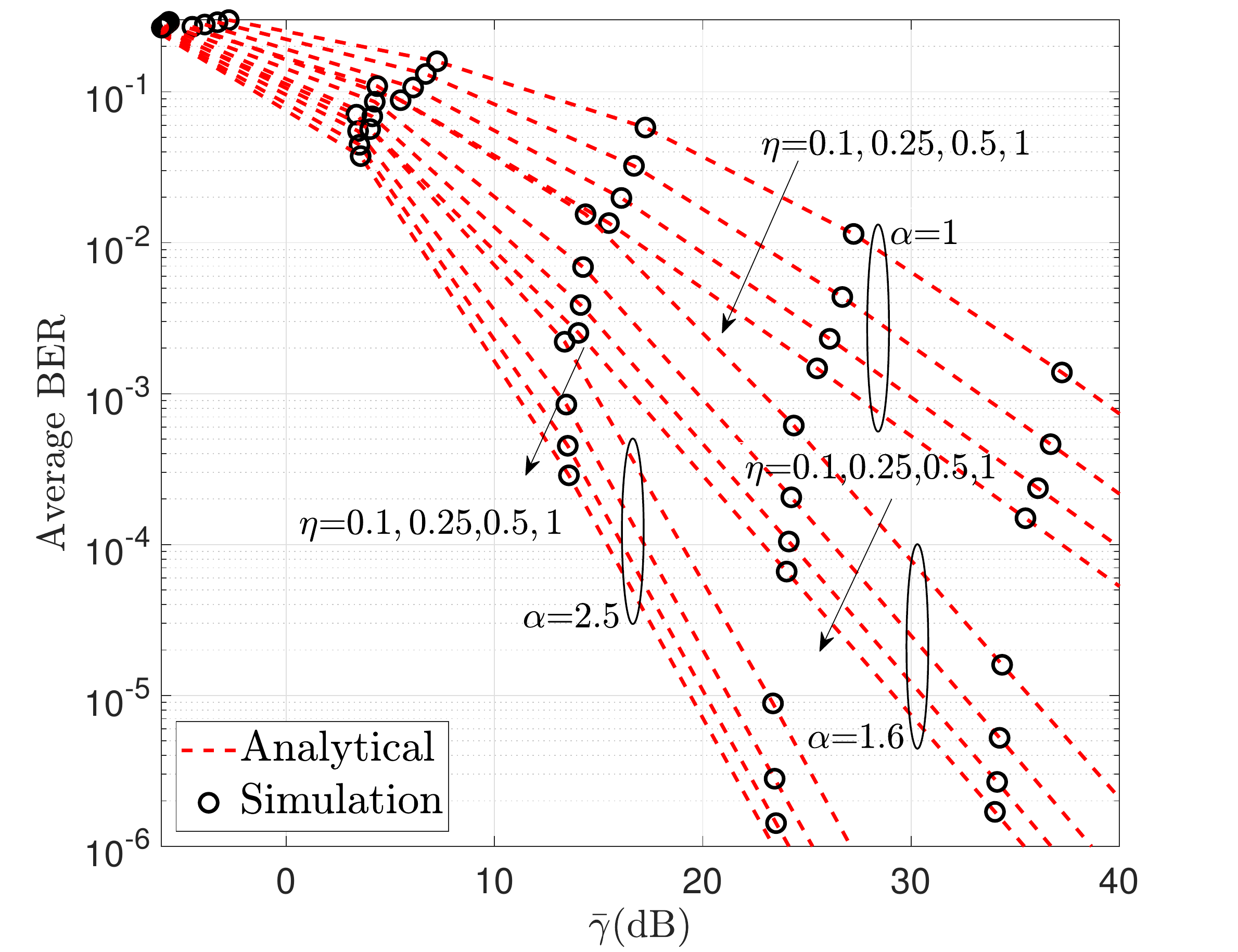}} 
	\caption{Outage probability and average BER performance of a wireless link over.}
	\label{fig:outage_ber}	
\end{figure*}
\section{Simulation and Numerical Results}
In this section, we use numerical and simulation analysis to demonstrate the superiority of the developed statistical results compared with the existing infinite series representation. We also verify the derived analytical results (using the Fox's H-function code of \cite{Alhennawi2016}) with Monte-Carlo simulations. To simulate random samples of the envelope $R$ in \eqref{eq:Ralpha}  quantified by $\alpha$, $\eta$, $\kappa$, $\mu$, $p$, $q$, and $\hat{r}$, we use the following parametric equations \cite{Yacoub_2016_alpha_eta_kappa_mu}:
\small
\begin{flalign}
&\sigma_x = \sqrt{\frac{\eta (p+1)\hat{r}^\alpha}{2(\eta+1)(\kappa+1)\mu p}}, ~~ \sigma_y = \sqrt{\frac{ (p+1)\hat{r}^\alpha}{2(\eta+1)(\kappa+1)\mu p}}\nonumber \\ & 	\lambda_x = \sqrt{\frac{\eta \kappa q \hat{r}^\alpha}{(q\eta+1)(\kappa+1)}}, ~~~~~\lambda_y = \sqrt{\frac{ \kappa \hat{r}^\alpha}{(q\eta+1)(\kappa+1)}} \nonumber \\ &  \mu_x = \frac{2p\mu}{1+p}, ~~~~~~~~~~~~~~~~~~~~~  \mu_y = \frac{2\mu}{1+p}. \nonumber
\end{flalign}
\normalsize
In Fig. \ref{fig:pdf_cdf}, we numerically evaluate the derived PDF and CDF of the channel envelope using the software package of multivariate Fox's H-function \cite{Alhennawi2016}. We take $\alpha=2$, $\eta=1$, $\mu=2$, $\kappa=1$, $p=3$, $q=1$, $\hat{r}=1$. We validate the derived statistical results (as given in \eqref{eq:pdf_new_without_pe} and \eqref{eq:cdf_new_without_pe}) by Monte Carlo simulations obtained through random samples of the envelope $R$ and MATLAB function \emph{histogram ([R], 'Normalization', 'pdf')}, where $[R]$ denotes an array of $10^7$ samples of the envelope $R$. We also compare the derived analytical results with the state-of-the-art by numerically evaluating the infinite series  PDF in  \cite[eq. 11]{Silva_2020_alpha_eta_kappa_mu} (as represented in \eqref{eq:pdf_series}) and the infinite series CDF in \cite[eq. 12]{Silva_2020_alpha_eta_kappa_mu} for various number of summands in \eqref{eq:pdf_series}. 

Fig. \ref{fig:pdf_cdf}(a) shows that the infinite-series PDF underestimates the actual PDF when the number of summands are small ($n=2$  and $n=5$) for given system parameters. Even for higher iteration $n=15$, there is an error with the infinite-series PDF in the tail region of the distribution function $x\geq 1.5$. Of course, the infinite-series is convergent, achieving the exact within $20$ summands. Note that the number of summands for the convergence of the infinite-series may vary according to the change in the channel parameters for different environmental conditions approximating the system design from the actual if a sufficient number of summands are not considered. Moreover, Fig. 1(a) depicts  that the asymptotic PDF of \eqref{eq:asympt_pdf} is close to exact values when $r \to 0$, similar to the asymptotic PDF of \cite[eq.19]{Silva_2020_alpha_eta_kappa_mu}. However, the derived asymptotic PDF also approximates the shape of the actual PDF, unlike the asymptotic PDF of \cite[eq.19]{Silva_2020_alpha_eta_kappa_mu}.
 In Fig. \ref{fig:pdf_cdf}(b), we demonstrate the significance of the derived  CDF of \eqref{eq:cdf_new_without_pe} compared with the infinite-series CDF of \cite{Silva_2020_alpha_eta_kappa_mu} depicting similar observations and conclusions, as illustrated through Fig. \ref{fig:pdf_cdf} (a).

In Fig. \ref{fig:outage_ber}, we demonstrate the performance of a $50$ \mbox{m} wireless link by computing the average SNR $\bar{\gamma}$ at a carrier frequency of $275$ \mbox{GHz}, antenna gains of $40$ \mbox{dBi}, and atmospheric absorption coefficient calculated from Table II and Table III from \cite{Pranay_2021_TVT} for a range of transmit power $-40$\mbox{dBm} to $0$\mbox{dBm}. Fig. \ref{fig:outage_ber}(a) depicts the outage performance of the system for different values of clustering parameter $\mu$ and the parameter $\kappa$. The outage performance of the system improves with an increase in the parameter $\mu$, signifying an increase in the number of multi-path clusters. The outage probability also decreases with an increase in $ \kappa $ due to the improvement in the power of the dominant component compared with the scattered waves. Fig. \ref{fig:outage_ber}(a) shows that the outage probability decreases from $2.6\times 10^{-3}$ to $1.3\times10^{-5}$ when  $ \mu $ is increased from $1$ to $2$ with $ \kappa=0.2 $ at $30$ \mbox{dB} of average SNR. However, the improvement in the outage probability with an increase in the parameter $\kappa$ is not significant: the outage probability  decreases from $1.3\times 10^{-5}$ to $1.5\times10^{-6}$ when $\kappa$ is increased from $0.2$ to $2$ with $\mu=2$  at $30$ \mbox{dB} average SNR. Further, the derived asymptotic results for outage probability match very closely with analytical and simulated results at a reasonably high SNR. It should be emphasized that the slope of the outage probability changes with $\mu$ and remains constant with $\kappa$ verifying the diversity order (i.e., $\frac{\alpha\mu}{2}$) for the $\alpha$-$\eta$-$\kappa$-$\mu$ fading channel.

In Fig. \ref{fig:outage_ber} (b), we illustrate the average BER performance of the system by varying $\alpha$ and $\eta$ parameters for the BPSK modulation $\{ p_m  = 0.5, q_m  = 1\}$. The figure shows that the average BER decreases with an increase in $\alpha$ since the wireless channel becomes more linear. Moreover, an increase in the value of $\eta$ reduces the average BER, as demonstrated in the existing literature. The figure shows that the average BER improves by $10$ times when the value of $\eta$ is increased from $0.1$ to $1$ with $\alpha=1$ at $30$ \mbox{dB} of average SNR. However, the non-linearity factor $\alpha$ changes the average BER more significantly: the average BER reduces from $6.5\times10^{-3}$ to $8\times10^{-5}$ (a factor $80$ reduction) when $\alpha =1$ (highly non-linear channel) increases to $\alpha =2.5$ (less non-linear channel) with $\eta=0.1$ at  $30$ \mbox{dB} of average SNR.

\section{Conclusions}
We derived exact PDF and CDF for the envelope of the $\alpha$-$\eta$-$\kappa$-$\mu$ fading model using a single Fox's H-function without containing any infinite-series approximation for an accurate performance assessment of wireless systems.   To substantiate the superiority of the developed analysis,   we presented the exact outage probability and average BER of a wireless  link subjected to the $\alpha$-$\eta$-$\kappa$-$\mu$ fading. The exact analytical expressions provide an elegant framework to develop  asymptotic analysis  in a high SNR region using simpler Gamma functions.  We verified the accuracy of  derived analytical results by comparing them with  Monte Carlo simulations. 

We envision that the proposed statistical analysis on the envelope may rekindle the interest in the $\alpha$-$\eta$-$\kappa$-$\mu$ model attributed to its excellent fitting to the experimental data and now mathematically tractable for performance analysis in the next-generation wireless systems.

\section*{Appendix A}
Defining $U = \sum_{i=1}^{\mu_x} (X_i+\lambda_{x_i})^2$ and $V = \sum_{i=1}^{\mu_y} (X_i+\lambda_{y_i})^2$, i.e. $R^\alpha = U+V$ . By means of transformation of variables, the PDF $f_R(r)$ of $ R $ is found as \cite{Silva_2020_alpha_eta_kappa_mu}
\begin{flalign} \label{eq:derivation_1}
f_R(r) = \alpha r^{\alpha-1} \int_{0}^{r^\alpha} f_{U}(r^\alpha-v) f_{V}(v) dv
\end{flalign}
Using $ f_U (u) $ and $ f_V (v) $ from \cite{Silva_2020_alpha_eta_kappa_mu} in \eqref{eq:derivation_1} and after some algebraic manipulation, we get\footnote{There is a typo in equation (9) of \cite{Silva_2020_alpha_eta_kappa_mu}. It should be $(\hat{r}^\alpha)^{1+\frac{\mu}{2}}$ in the denominator of the second term of the first row.}
\begin{flalign} \label{eq:pdf_eq9}
f_{R}(r) &= \frac{\psi_1 r^{\psi_2}}{(\hat{r}^\alpha)^{1+\frac{\mu}{2}}} \times e^{-\psi_3 r^{\alpha}} \int_{0}^{r^\alpha}
\frac{(r^{\alpha}-v)^{\frac{A_1}{2}}}{v^{-\frac{A_2}{2}}}  e^{-A_{3}v} \nonumber \\ & \times
I_{A_{1}}(A_{4}(r^\alpha-v)^{\frac{1}{2}}) I_{A_{2}}(A_{5}v^{\frac{1}{2}})dv
\end{flalign}
where $I_{A_{1}}$ and $I_{A_{2}}$ are the modified Bessel function of the first kind. Using Meijer's G representation of the Bessel functions \cite{Mathematica_besseli}, \eqref{eq:pdf_eq9} can be rewritten as
\begin{flalign} \label{eq:pdf_derive_1}
f_{R}(r)&=  \frac{\psi_1 r^{\psi_2}}{(\hat{r}^\alpha)^{1+\frac{\mu}{2}}}  e^{-\psi_3r^\alpha} \pi 2^{-A_1} (A_{4}(r^\alpha-v)^{\frac{1}{2}})^{A_1} \nonumber\\ & \times  \pi 2^{-A_2} (A_{5}v^{\frac{1}{2}})^{A_2}  \int_{0}^{r^\alpha}\frac{(r^{\alpha}-v)^{\frac{A_1}{2}}}{v^{-\frac{A_2}{2}}}  G_{0,1}^{1,0}\left(\begin{array}{c} -\\0 \end{array}\left| A_3 v\right. \right) \nonumber \\ & \times G_{1,3}^{1,0} \left(\begin{array}{c}\frac{1}{2}\\0, -A_1, \frac{1}{2} \end{array} \left | \frac{A_{4}^2(r^\alpha -v)}{4}\right. \right)  \nonumber\\ & \times   G_{1,3}^{1,0} \left (\begin{array}{c} \frac{1}{2}\\0, -A_2, \frac{1}{2} \end{array} \left| \frac{A_{5}^2v}{4}\right. \right) dv
\end{flalign}
Utilizing the integral representation of Meijer's G-function \cite{Mathai_2010}, we can represent \eqref{eq:pdf_derive_1} as 
\small
\begin{flalign} \label{eq:pdf_derive_3}
&f_{R}(r) = \frac{\psi_1}{(\hat{r}^\alpha)^{1+\frac{\mu}{2}}} \pi^2 2^{(2-\mu)} A_4^{A_1} A_5^{A_2} r^{\psi_2} e^{-\psi_3r^\alpha} \nonumber \\ & \times \frac{1}{(2\pi i)^{3}} \int_{\mathcal{L}_1}\int_{\mathcal{L}_2}\int_{\mathcal{L}_3}\Gamma(-s_1){A_3}^{s_1} ds_1 \nonumber \\ & \times \frac{\Gamma(-s_2)}{\Gamma(1+A_1+s_2)\Gamma(\frac{1}{2}+s_2)\Gamma(\frac{1}{2}-s_2)}\Big(\frac{A_{4}^{2}}{4} \Big)^{s_2} ds_2 \nonumber
\\& \times \frac{\Gamma(-s_3)}{\Gamma(1+A_2+s_3)\Gamma(\frac{1}{2}+s_3)\Gamma(\frac{1}{2}-s_3)}\Big(\frac{A_{5}^{2}}{4} \Big)^{s_3} ds_3 \times I_1
\end{flalign}
\normalsize
where $\mathcal{L}_1$, $\mathcal{L}_2$, and $\mathcal{L}_3$, denote the contour integrals. The inner integral $I_1 $ can be represented and simplified using the identity \cite[3.191.1]{Gradshteyn} as
\begin{flalign} \label{inn}
&I_1 = \int_{0}^{r^\alpha} {v}^{{A_2}+s_1+s_3} (r^{\alpha}-v)^{{A_1}+s_2} dv  = \nonumber \\  & \frac{\Gamma(1+A_2+s_1+s_3)\Gamma(1+A_1+s_2)}{\Gamma(2+A_2+s_1+s_3+A_1+s_2)}  
{r^{\alpha(1+A_2+s_1+s_3+A_1+s_2)}}
\end{flalign}

Finally, substituting  \eqref{inn} in \eqref{eq:pdf_derive_3}, rearranging the terms, and applying the definition of multivariate Fox's H-function \cite{Mathai_2010}, we get the PDF of Theorem 1 in \eqref{eq:pdf_new_without_pe}, which  concludes the proof. 

\section*{Appendix B}
The CDF $F_{R}(r)$ can be derived by integrating the PDF of \eqref{eq:pdf_new_without_pe} as $\int_{0}^{r} f_{R}(r) dr $ together with the definition of Fox's H-function:
\begin{flalign} \label{eq:cdf_derive_1}
&F_R(r) = \frac{\psi_1}{(\hat{r}^\alpha)^{1+\frac{\mu}{2}}} \pi^2 2^{(2-\mu)} A_4^{A_1} A_5^{A_2} \frac{1}{(2\pi i)^{3}} \nonumber \\ & \times  \int_{\mathcal{L}_1}\int_{\mathcal{L}_2}\int_{\mathcal{L}_3} \frac{\Gamma(-s_1) \Gamma(-s_2)\Gamma(1+{A_1}+s_2)}{\Gamma(1+A_1+s_2)\Gamma(\frac{1}{2}+s_2)\Gamma(\frac{1}{2}-s_2)}
\nonumber \\ &  \times \frac{\Gamma(-s_3)}{\Gamma(1+A_2+s_3)\Gamma(1/2+s_3)\Gamma(1/2-s_3)} \nonumber \\ & \times
\frac{\Gamma(1+{A_2}+s_1+s_3)}{\Gamma(2+{A_2}+s_1+s_3+{A_1}+s_2)} \nonumber \\ & \times
(A_{3}r^\alpha)^{s_1} \Big(\frac{A_{4}^{2}}{4}r^\alpha\Big)^{s_2} \Big(\frac{A_{5}^{2}}{4}r^\alpha\Big)^{s_3} ds_1 ds_2 ds_3  \nonumber \\ & \times \int_{0}^{r}  r^{\alpha\mu-1}  e^{-\psi_3 r^\alpha}  dr
\end{flalign}
To solve the inner integral, we express the exponential term using Mellin's transform as:
\begin{flalign}\label{mellin}
e^{-\psi_3 r^\alpha}= \frac{1}{2\pi i} \int_{\mathcal{L}_4} \Gamma(0-s_4) (\psi_3 r^\alpha)^{s_4} ds_4
\end{flalign}
Using \eqref{mellin} in  \eqref{eq:cdf_derive_1}, we express

\begin{flalign}\label{inner}
&\int_{0}^{r} r^{\alpha(s_1+s_2+s_3+s_4)+{\alpha\mu}-1}d\gamma  = \nonumber \\ &  r^{\alpha(s_1+s_2+s_3+s_4)+{\alpha\mu}}  \frac{\Gamma(\alpha(s_1+s_2+s_3+s_4)+\alpha\mu)}{\Gamma(\alpha(s_1+s_2+s_3+s_4)+\alpha\mu+1)}
\end{flalign}
Finally, using \eqref{inner} in  \eqref{eq:cdf_derive_1} with some simplification and applying the definition of multivariate Fox's H-function \cite{Mathai_2010}, we get the CDF of Lemma 1 in \eqref{eq:cdf_new_without_pe} to conclude the proof.

\bibliographystyle{IEEEtran}
\bibliography{alpha_eta_kappa_mu}

\begin{thebibliography}{10}
\providecommand{\url}[1]{#1}
\csname url@samestyle\endcsname
\providecommand{\newblock}{\relax}
\providecommand{\bibinfo}[2]{#2}
\providecommand{\BIBentrySTDinterwordspacing}{\spaceskip=0pt\relax}
\providecommand{\BIBentryALTinterwordstretchfactor}{4}
\providecommand{\BIBentryALTinterwordspacing}{\spaceskip=\fontdimen2\font plus
\BIBentryALTinterwordstretchfactor\fontdimen3\font minus
  \fontdimen4\font\relax}
\providecommand{\BIBforeignlanguage}[2]{{%
\expandafter\ifx\csname l@#1\endcsname\relax
\typeout{** WARNING: IEEEtran.bst: No hyphenation pattern has been}%
\typeout{** loaded for the language `#1'. Using the pattern for}%
\typeout{** the default language instead.}%
\else
\language=\csname l@#1\endcsname
\fi
#2}}
\providecommand{\BIBdecl}{\relax}
\BIBdecl

\bibitem{Dang_2020_nature}
S.~Dang \emph{et~al.}, ``What should {6G} be?'' \emph{Nature Electron}, no.~3,
  p. 20–29, 2020.

\bibitem{Akyildiz_2020}
I.~F. Akyildiz \emph{et~al.}, ``{6G} and beyond: The future of wireless
  communications systems,'' \emph{IEEE Access}, vol.~8, pp. 133\,995--134\,030,
  2020.

\bibitem{Serghiou_2022_THz_survey}
D.~Serghiou \emph{et~al.}, ``Terahertz channel propagation phenomena,
  measurement techniques and modeling for {6G} wireless communication
  applications: a survey, open challenges and future research directions,''
  \emph{IEEE Commun. Surv. Tut.}, pp. 1--1, 2022.

\bibitem{Thomas1994}
H.~Thomas \emph{et~al.}, ``An experimental study of the propagation of 55 {GHz}
  millimeter waves in an urban mobile radio environment,'' \emph{IEEE Trans.
  Veh. Technol.}, vol.~43, no.~1, pp. 140--146, 1994.

\bibitem{Smulders2009}
P.~F.~M. Smulders, ``Statistical characterization of {60-GHz} indoor radio
  channels,'' \emph{IEEE Trans. Ant. Propag.}, vol.~57, no.~10, pp. 2820--2829,
  2009.

\bibitem{Moon2005}
M.-S. Choi, Grosskopf, and Rohde, ``Statistical characteristics of 60 {GHz}
  wideband indoor propagation channel,'' in \emph{2005 IEEE 16th Int. Symp. on
  Personal, Indoor and Mobile Radio Commun.}, vol.~1, 2005, pp. 599--603.

\bibitem{Samimi2016}
M.~K. Samimi \emph{et~al.}, ``28 {GHz} millimeter-wave ultrawideband
  small-scale fading models in wireless channels,'' in \emph{2016 IEEE 83rd
  Veh. Technol. Conf. (VTC Spring)}, 2016, pp. 1--6.

\bibitem{Romero2016}
J.~M. Romero-Jerez \emph{et~al.}, ``The fluctuating two-ray fading model for
  mmwave communications,'' in \emph{2016 IEEE Globecom Workshops (GC Wkshps)},
  2016, pp. 1--6.

\bibitem{Marins2019_alpha_eta_kapp_mu}
T.~R.~R. Marins \emph{et~al.}, ``Fading evaluation in the {mm-Wave} band,''
  \emph{IEEE Trans. Commun.}, vol.~67, no.~12, pp. 8725--8738, 2019.

\bibitem{Papasotiriou2021_scientific_report}
E.~Papasotiriou \emph{et~al.}, ``An experimentally validated fading model for
  {THz} wireless systems,'' \emph{Scientific report}, vol.~11, 2021.

\bibitem{Du2022_FTR_THz_RIS}
H.~Du \emph{et~al.}, ``Performance and optimization of reconfigurable
  intelligent surface aided {THz} communications,'' \emph{IEEE Trans. Commun.},
  vol.~70, no.~5, pp. 3575--3593, 2022.

\bibitem{Yacoub_2016_alpha_eta_kappa_mu}
M.~D. Yacoub, ``The $\alpha $ - $\eta $ - $\kappa $ - $\mu $ fading model,''
  \emph{IEEE Trans. Ant. Prop.}, vol.~64, no.~8, pp. 3597--3610, 2016.

\bibitem{Silva_2020_alpha_eta_kappa_mu}
C.~R.~N. da~Silva \emph{et~al.}, ``The $\alpha$ - $\eta$ - $\kappa$ - $\mu$
  fading model: New fundamental results,'' \emph{IEEE Trans. Ant. Prop.},
  vol.~68, no.~1, pp. 443--454, 2020.

\bibitem{Li_2017_alpha_eta_kappa_mu}
X.~Li \emph{et~al.}, ``Capacity analysis of $\alpha $ - $\eta $ - $\kappa $ -
  $\mu $ fading channels,'' \emph{IEEE Commun. Lett.}, vol.~21, no.~6, pp.
  1449--1452, 2017.

\bibitem{Mathur_2018_alpha_eta_kappa_mu}
A.~Mathur \emph{et~al.}, ``On physical layer security of $\alpha$ - $\eta$ -
  $\kappa$ - $\mu$ fading channels,'' \emph{IEEE Commun. Lett.}, vol.~22,
  no.~10, pp. 2168--2171, 2018.

\bibitem{Jia2018_alpha_eta_kapp_mu}
S.~Jia \emph{et~al.}, ``Performance analysis of physical layer security over
  $\alpha$ – $\eta$ – $\kappa$ – $\mu$ fading channels,'' \emph{China
  Commun.}, vol.~15, no.~11, pp. 138--148, 2018.

\bibitem{Moualeu_2019_alpha_eta_kappa_mu}
J.~M. Moualeu \emph{et~al.}, ``On the performance of $\alpha$ – $\eta$ –
  $\kappa$ – $\mu$ fading channels,'' \emph{IEEE Commun. Lett.}, vol.~23,
  no.~6, pp. 967--970, 2019.

\bibitem{Singh2019_alpha_eta_kapp_mu}
R.~Singh and M.~Rawat, ``On the analysis of effective capacity for {5G}
  millimeter-wave communication,'' in \emph{2019 10th Int. Conf. Comput.,
  Commun. and Netw. Technol. (ICCCNT)}, 2019, pp. 1--4.

\bibitem{Ai_2020_alpha_eta_kappa_mu}
Y.~o. Ai, ``Effective throughput analysis of $\alpha $ - $\eta $ - $\kappa $ -
  $\mu $ fading channels,'' \emph{IEEE Access}, vol.~8, pp. 57\,363--57\,371,
  2020.

\bibitem{Al2020_alpha_eta_kapp_mu}
H.~Al-Hmood and H.~S. Al-Raweshidy, ``On the effective rate and energy
  detection based spectrum sensing over $\alpha -\eta -\kappa -\mu$ fading
  channels,'' \emph{IEEE Trans. Veh. Technol.}, vol.~69, no.~8, pp. 9112--9116,
  2020.

\bibitem{Saraereh2020_alpha_eta_kapp_mu}
O.~A. Saraereh \emph{et~al.}, ``Interference analysis for vehicle-to-vehicle
  communication at 28 {GHz},'' \emph{Electronics}, vol.~9, no.~2, 2020.

\bibitem{Soni2020_alpha_eta_kapp_mu}
B.~Soni \emph{et~al.}, ``Performance analysis of {NOMA} aided cooperative
  relaying over $\alpha -\eta -\kappa -\mu$ fading channels,'' in \emph{2020
  Nat. Conf. Commun. (NCC)}, 2020, pp. 1--6.

\bibitem{Vishwakarma2021_alpha_eta_kapp_mu}
N.~Vishwakarma and {R. Swaminathan}, ``Performance analysis of hybrid {FSO/RF}
  communication over generalized fading models,'' \emph{Opt. Commun.}, vol.
  487, p. 126796, 2021.

\bibitem{GOSWAMI2019_alpha_eta_kapp_mu}
A.~Goswami and A.~Kumar, ``Performance analysis of multi-hop wireless
  communication systems over $\alpha -\eta -\kappa -\mu$ channel,''
  \emph{Physical Commun.}, vol.~33, pp. 9--15, 2019.

\bibitem{Kavaiya2020_alpha_eta_kapp_mu}
S.~Kavaiya \emph{et~al.}, ``On physical layer security over $\alpha$ – $\eta$
  – $\kappa$ – $\mu$ fading for relay based vehicular networks,'' in
  \emph{2020 Int. Conf. Signal Process. and Commun. (SPCOM)}, 2020, pp. 1--5.

\bibitem{Mathai_2010}
A.~M.{ Mathai} \emph{et~al.}, \emph{The {H}-Function: Theory and
  Applications}.\hskip 1em plus 0.5em minus 0.4em\relax Springer, New York, NY,
  2010.

\bibitem{Meijers_asymp_low_snr}
\emph{The Wolfram function Site}, Available:
  https://functions.wolfram.com/HypergeometricFunctions/MeijerG/ 06/
  01/05/01/0006/ (Accessed: April 11, 2023).

\bibitem{Ansari2011}
I.~S. {Ansari} \emph{et~al.}, ``A new formula for the {BER} of binary
  modulations with dual-branch selection over generalized-{K} composite fading
  channels,'' \emph{IEEE Trans. Commun.}, vol.~59, no.~10, pp. 2654--2658,
  2011.

\bibitem{Gradshteyn}
I.~S. {Gradshteyn} and I.~M. {Ryzhik }, \emph{Table of Integrals, Series, and
  Products}.\hskip 1em plus 0.5em minus 0.4em\relax Academic press, San Diego,
  CA, 6th edition, 2000.

\bibitem{Alhennawi2016}
H.~R. Alhennawi \emph{et~al.}, ``Closed-form exact and asymptotic expressions
  for the symbol error rate and capacity of the {$H$}-function fading
  channel,'' \emph{IEEE Trans. Veh. Technol.}, vol.~65, no.~4, pp. 1957--1974,
  2016.

\bibitem{Pranay_2021_TVT}
P.~Bhardwaj and S.~M. Zafaruddin, ``Performance of dual-hop relaying for
  {THz-RF} wireless link over asymmetrical $\alpha$-$\mu$ fading,'' \emph{IEEE
  Trans. Veh. Technol.}, vol.~70, no.~10, pp. 10\,031--10\,047, 2021.

\bibitem{Mathematica_besseli}
\emph{The Wolfram function Site}, Available:
  https://functions.wolfram.com/Bessel-TypeFunctions/BesselY/
  introductions/Bessels/ (Accessed: Sept. 26, 2022).

\end{thebibliography}
\end{document}